\documentclass[conference]{IEEEtran}
\usepackage{array,cite}
\usepackage{graphicx,epsfig,subfigure}
\usepackage{amsthm}
\usepackage{amsmath}
\usepackage{mdwmath}
\usepackage{array}
\usepackage{subeqnarray}
\usepackage{stfloats}
\usepackage{algorithm,algorithmic}
\usepackage{xcolor}
\usepackage{amssymb}

\newtheorem{theorem}{Theorem}
\newtheorem{lemma}{Lemma}
\newtheorem{definition}{Definition}
\newtheorem{corollary}{Corollary}

\begin{document}

\title{Impact of Mobility on the Sum Rate of NB-OFDMA Based Mobile IoT Networks}

\author{\IEEEauthorblockN{Chunxu~Jiao, Zhaoyang~Zhang$^{\dagger}$, and Caijun~Zhong}
\IEEEauthorblockA{Department of Information Science and Electronic Engineering, Zhejiang University, Hangzhou 310027, China\\
E-mail: \{jiaocx1990, ning\_ming$^{\dagger}$, caijunzhong\}@zju.edu.cn}}

\maketitle

\begin{abstract}
In future Internet of Things (IoT) networks, the explosive growth of mobile devices compel us to reconsider the effectiveness of the current frequency-division multiple access (FDMA) schemes. Devices' differentiated mobility features and diversified scattering environments make it more complicated to characterize the multi-user interference. In this paper, we thoroughly analyze the impacts of devices' mobility on the inter-sub-carrier interference (ICI) in an IoT system based on the 3GPP narrow-band orthogonal frequency-division multiple access (NB-OFDMA) protocol, and obtain the relationship between the system sum-rate and devices' mobility. Our results may shed some lights on the system design under the mobile scenarios.
\end{abstract}

\section{Introduction}

% Motivation: background and application-drives problems we are facing, how tough is the problem
In recent years, Internet of Things (IoT) has received a lot of attention in research and practice. It has been adopted in various crucial applications, such as smart logistics, smart home, smart healthcare and intelligent vehicles \cite{gubbi2013internet,atzori2011siot}. As predicted by Cisco, there will be over $50$ billion smart things connected to the Internet by $2020$ \cite{evans2011internet}. Such multitudinous IoT devices are bound to consume a considerable amount of resources and result in new issues that must be taken care of.

% Existing Efforts: a comprehensive summary of related works, acknowledge their merits first and then point out their shortcomings
Since the future IoT will probably take the form of a cloud-based mobile network \cite{yau2014intelligent}, most of the IoT devices move around constantly, such as self-driving cars, high speed trains and mobile sensors. These multifarious mobile sensing and computing devices are expected to provide huge amounts of physical data to the IoT cloud. Given such highly dense networks, devices' mobility undeniably incur more severe multi-user interference and deteriorates the system throughput. In most of the previous papers, however, system sum-rate is analyzed only under quasi-static fading conditions \cite{wang2013uplink,hassibi2003much}, which is not practical in future mobile communications scenarios. In this sense, it becomes immensely important to take into account the performance degradation caused by devices' mobility. Although some researchers have investigated the impact of the maximum Doppler frequency on system performance \cite{robertson1999effects,li2001bounds}, they mainly concentrate on orthogonal frequency-division multiplexing (OFDM) settings. Some later works focus on orthogonal frequency-division multiple access (OFDMA) protocols, but the system impairments were incurred by frequency offsets \cite{huang2005interference} or the two-tier femtocell network architecture \cite{lopez2009ofdma}. Devices' mobility features and the induced multi-user interference have not been fully investigated.

In this paper, based on the newly proposed narrow band OFDMA (NB-OFDMA) \cite{3gpp2015mobile}, we thoroughly analyze the impacts of devices' mobility on the inter sub-carrier interference (ICI) and the system sum-rate in the uplink of a IoT cellular communication system. Devices are modeled as Possion point processes (PPP) and are characterized with independent velocities and moving directions. Under broadband multi-path fading conditions, the expression of the total ICI power suffered by each device is derived. The corresponding upper and lower bounds are also provided, from which an approximate relationship between the interference level and devices' maximum velocity is formulated. Subsequently, we analyze the performance of NB-OFDMA under various metrics of interest (e.g., single-device ergodic capacity and system sum-rate). Note that while we focus on the scenario of NB-OFDMA, the results in this paper may apply to the general OFDMA situations.

The rest of the paper is organized as follows. In Section II, the system model is described. In Section III, we theoretically analyze the impacts of devices' mobility on the inter sub-carrier interference, single-device ergodic capacity and system sum-rate. Section IV gives the relevant numerical results. And finally, Section V concludes the work.

\section{System Model}
In this section, the mobility features of moving devices and the multi-path effects of the propagation channel in cellular systems are described. The uplink transmission model is also presented.

\subsection{NB-OFDMA Cellular System}
\begin{figure}[!t]\centering
\includegraphics[angle=0,scale=0.23]{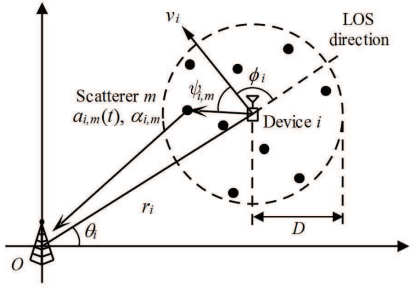}
\caption{The position and mobility of device $i$ are characterized by $\left( r_{i},\theta_{i} \right)$ and $\left( v_{i},\phi_{i} \right)$, respectively. And, scatterer $m$ possesses a time-varying attenuation $a_{i,m}\left(t\right)$, a random phase shift $\alpha_{i,m}$ and a delay $\tau_{i,m}$.}
\label{Fig_path_model}
\end{figure}

As depicted in Fig. \ref{Fig_path_model}, an NB-OFDMA based cellular system is considered. A base station (BS) comprising an omni-directional antenna is located at the center of the cell. $K$ single-antenna mobile devices are randomly distributed around the BS. The radius of the cell is $R$ and this NB-OFDMA system has access to $2N+1$ sub-carriers\footnote{For a conventional NB-OFDMA system, the number of sub-carriers is $2N$ in most cases\cite{3gpp2015mobile}. However, we use an odd number of sub-carriers to facilitate the ensuing analysis. Extension to $2N$ sub-carriers is straightforward.}. Under this setting, we restrict our attention on a fully loaded system, where each device is assigned with one single sub-carrier, i.e., $K=2N+1$.

For a given time instant, using polar coordinates, we can easily characterize the the position of device $i$ as $\left( r_{i},\theta_{i} \right)$. Assume that the instantaneous locations of mobile devices are determined through a series of Poisson point processes, then the probability density functions (PDF) of $r_{i}$ and $\theta_{i}$ are respectively given by
\begin{equation}
\begin{split}
p(r_{i})&=\left\{\begin{matrix}
\frac{2r_{i}}{R^{2}},~0\leq r_{i}\leq R
\\
0,~~~\textrm{otherwise}~~
\end{matrix}\right. \mbox{ and}\\
p(\theta_{i})&=\left\{\begin{matrix}
\frac{1}{2\pi},~0\leq \theta_{i}< 2\pi
\\
0,~~\textrm{otherwise}~~~
\end{matrix}\right..
\label{Location_distribution}
\end{split}
\end{equation}
Meanwhile, let $v_{i}$ denote the instantaneous velocity of device $i$ and $\phi_{i}$ be the included angle between the moving direction and the line-of-sight direction. Assume that devices have independent mobility features, then it is reasonable to assume that devices' velocities and moving directions are i.i.d. random variables with uniform distributions, i.e., their PDFs are
\begin{equation}
\begin{split}
p(v_{i})&=\left\{\begin{matrix}
\frac{1}{V_{\textrm{max}}},~0\leq v_{i}\leq V_{\textrm{max}}
\\
0,~~~~~\textrm{otherwise}~~~~~
\end{matrix}\right. \mbox{ and}\\
p(\phi_{i})&=\left\{\begin{matrix}
\frac{1}{2\pi},~0\leq \phi_{i}< 2\pi
\\
0,~~\textrm{otherwise}~~~
\end{matrix}\right.,
\label{Mobility_distribution}
\end{split}
\end{equation}
where $V_{\textrm{max}}$ denotes the maximum velocity of each device.

\subsection{Multi-path Channel Model}
In practice, the signal transmitted from device $i$ is reflected by numbers of surrounding local scatterers as shown in Fig. \ref{Fig_path_model}, and thereby propagating along several distinct paths. Assume that the waves transmitted from mobile devices are planar, and therefore, only single scattering occurs. In this sense, each scatterer contributes to a specific propagation path. Taking device $i$ for example, assume it is enclosed by $M_{i}$ local scatterers, then it is trivial that the coressponding multi-path channel $h_{i}\left(t\right)$ comprises $M_{i}$ paths. Without loss of generality, we suppose $M_{i}$ is known and fixed.

Next, again invoking PPP, these local scatterers are assumed to be uniformly distributed within a radius of $D$. Among them, scatterer $m$ (i.e., the $m$-th path) introduces a time-varying amplitude attenuation $a_{i,m}\left(t\right)$, a random phase shift $\alpha_{i,m}$ and a delay $\tau_{i,m}$. Then the multi-path channel can be modelled as
\begin{equation}
h_{i}\left(t\right)=\sum_{m=0}^{M_{i}-1}a_{i,m}\left(t\right)e^{j\alpha_{i,m}}\delta\left(t-\tau_{i,m}\right),
\label{Multi-path_Channel}
\end{equation}
where $\sum^{M_{i}-1}_{m=0}\mathbb{E}[ \left|a_{i,m}\left(t\right)\right|^{2} ]= 1$ since the channel is of unit power in the long term. Moreover, let $\psi_{i,m}$ denote the included angle between the $m$-th path and the moving direction of device $i$, which is uniformly distributed in $\left[0,2\pi \right)$. Under this setting, the Doppler frequency induced by the $m$-th path is obtained as
\begin{equation}
f_d^{(i,m)}=\frac{v_{i}}{\textrm{c}}f_{\textrm{c}}\cos( \psi_{i,m} ),
\label{Doppler}
\end{equation}
where $\textrm{c}$ is the wave propagation speed, $f_{\textrm{c}}$ denotes the carrier center frequency.

\subsection{Uplink Transmission Model}
Based on the modeling framework above, now we are able to derive the received signal at the BS as
\begin{align}
y\left(t\right)&=\sum_{j=-N}^{N}\sqrt{c r_{j}^{-\beta}P_{\textrm{T},j}}\sum_{m=0}^{M_{j}-1}\left(a_{j,m}\left(t\right)e^{j\alpha_{j,m}}\delta\left(t-\tau_{j,m}\right)\right) \notag\\
               &\ast \left(x_{j}\left(t\right)e^{j2\pi f_{d}^{(j,m)}t}\right)+n\left(t\right),
\label{System_model}
\end{align}
where
\begin{itemize}
    \item $c r_{j}^{-\beta}$ represents the path loss. $\beta$ is the path loss exponent; $c$ is the median of the mean path loss at the reference distance. $P_{\textrm{T},j}$ denotes the transmit power of device $j$.
    \item $x_{j}\left(t\right)=x_{j}e^{j2\pi f_{j} t}$ is the transmitted signal of device $j$. $n\left(t\right)$ is the complex Gaussian noise process.
\end{itemize}
We emphasize that, to mitigate the near-far effect, a power control scheme is invoked so that the average power levels of all signals at the BS are identical \cite{dai2006distributed}, i.e.,
\begin{equation}
c r_{i}^{-\beta} P_{\textrm{T},i}=c r_{j}^{-\beta} P_{\textrm{T},j},~\textrm{for}~i,j=-N,\dots,N.
\label{Effective_Power}
\end{equation}
Define $P_{\textrm{T}} \triangleq c r_{j}^{-\beta} P_{\textrm{T},j}$, then the path loss component in (\ref{System_model}) can be dropped. Furthermore, since the path gain $a_{j,m}\left(t\right)$ remains constant in one symbol period $T_{\textrm{s}}$, it is practical to discard the parameter $t$ and rewrite the transmission model as
\begin{align}
y\left(t\right)&=\sqrt{P_{\textrm{T}}}\sum_{j=-N}^{N}x_{j}\sum_{m=0}^{M_{j}-1}a_{j,m}e^{j\alpha_{j,m}}e^{j2\pi\left(f_{j}+f_{d}^{(j,m)} \right )\left(t-\tau_{j,m} \right )}\notag\\
               &+n\left(t\right).
\label{Effective_System_Model}
\end{align}
Now invoking sub-carrier $i$, the demodulated signal is
\begin{align}
y_{i}&=\sqrt{P_{\textrm{T}}}x_{i}\sum_{m=0}^{M_{i}-1}\frac{a_{i,m}e^{j\alpha_{i,m}}}{T_{\textrm{s}}}\int_{-\frac{T_{\textrm{s}}}{2}}^{\frac{T_{\textrm{s}}}{2}}e^{j2\pi f_{d}^{(i,m)}\left(t-\tau_{i,m}\right)}dt\notag\\
     &+\sum_{\substack{j=-N\\j\neq i}}^{N}\sum_{m=0}^{M_{j}-1}u_{i,j,m} + n_{i},
\label{Demodulated_Signal}
\end{align}
where
\begin{equation}
u_{i,j,m}\triangleq \frac{\sqrt{P_{\textrm{T}}}x_{j}a_{j,m}e^{j\alpha_{j,m}}}{T_{\textrm{s}}}\int_{-\frac{T_{\textrm{s}}}{2}}^{\frac{T_{\textrm{s}}}{2}}e^{j2\pi \left( f_{j}-f_{i}+f_{d}^{(j,m)}\right )\left(t-\tau_{j,m}\right)}dt
\label{individual-ICI}
\end{equation}
denotes the individual ICI component from device $j$ to device $i$ through its $m$-th path. Intuitively, the first term in (\ref{Demodulated_Signal}) represents the useful signal, the second one acts as the interference counterpart and $n_{i}$ is the the complex additive white Gaussian noise with mean zero and variance $0.5\sigma_{\textrm{n}}^{2}$ per dimension.

\begin{figure*}[!t]\centering
\includegraphics[angle=0,scale=0.18]{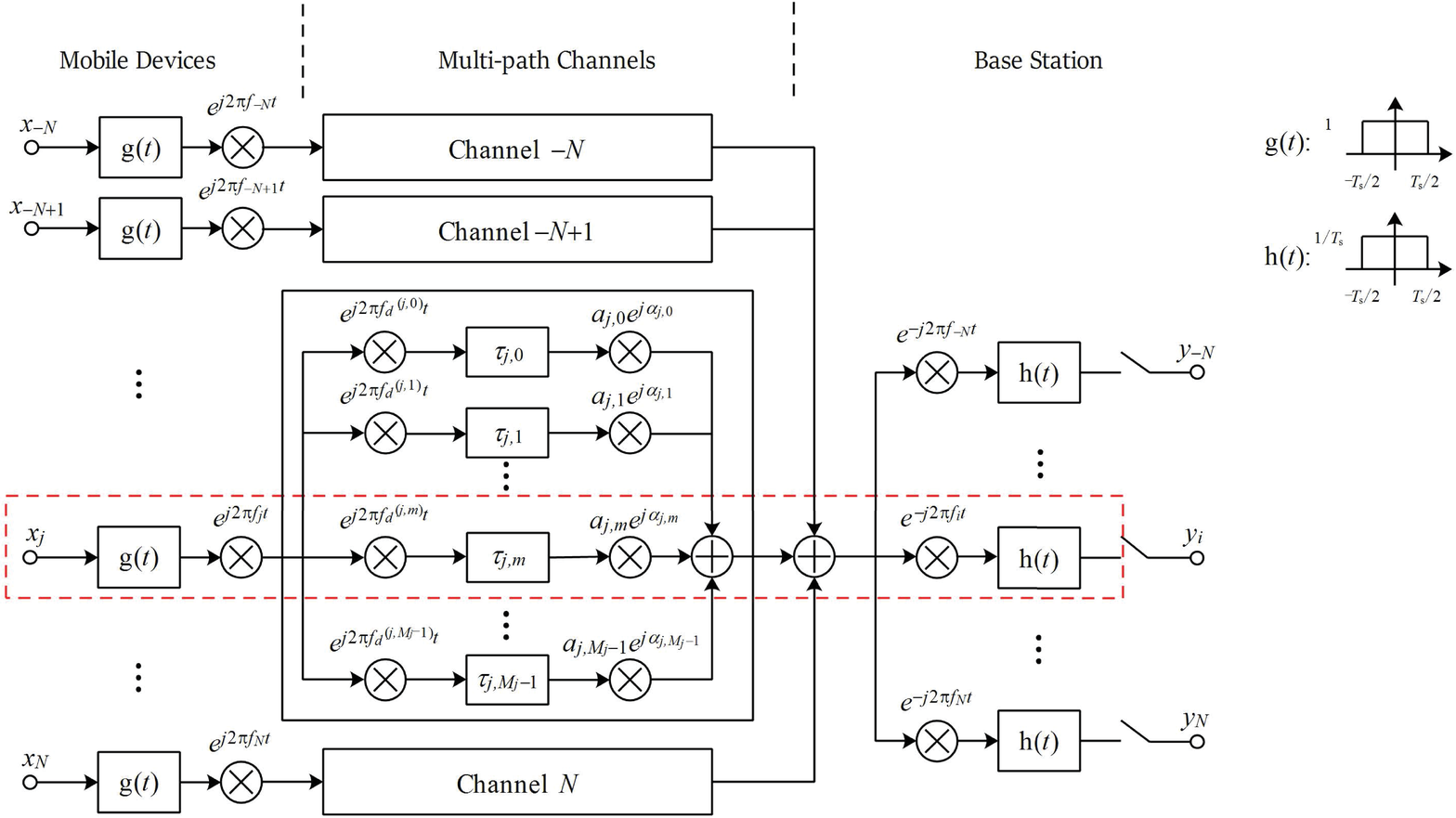}
\caption{The $m$-th path of device $j$ to BS is characterized by a specific Doppler frequency $f_d^{(j,m)}$, attenuation $a_{j,m}$, phase shift $\alpha^{(j,m)}$ and delay $\tau_{j,m}$.}
\label{Fig_data_flow_model}
\end{figure*}

\section{Interference Analysis}
In this section, we first construct the relation between ICI level and devices' mobility. Then, on this basis, the induced NB-OFDMA ergodic capacity and system sum-rate are investigated.

\subsection{Individual ICI}

To justify (\ref{Multi-path_Channel}) and (\ref{Demodulated_Signal}) more conveniently, we use Fig. \ref{Fig_data_flow_model} to illustrate our NB-OFDMA system. The pulse-shaping filters $g(t)$, $h(t)$ are matched and possess rectangular forms with length $T_{\textrm{s}}$. The signal transmitted from each device passes through a peculiar multi-path channel and is finally received and demodulated by the base station.

Within this framework, let us look into the demodulated signal using sub-carrier $i$. The disturbing interference comes from the other $2N$ sub-carriers due to the loss of orthogonality. To further elucidate how the data symbol of device $j$ influence the device $i$, we divide the propagation channel $h_{j}\left(t,\tau\right)$ into $M_{j}$ particular transmission links, as depicted by the red dashed-line-box. Under this setting, the power of the individual ICI component $u_{i,j,m}$ in (\ref{individual-ICI}) can be derived as \cite{robertson1999effects}
\begin{equation}
\sigma_{i,j,m}^{2}=\left|a_{j,m}\right|^{2}\textrm{sinc}^{2}\left( \left( f_{j}-f_{i}+f_{d}^{(j,m)} \right)T_{\textrm{s}} \right)P_{\textrm{T}},
\label{individual-ICI-power}
\end{equation}
where $\textrm{sinc}\left( x \right)=\frac{\sin\left( \pi x \right)}{\pi x}$. Due to the statistically independent data symbols $x_{j}$ and path gains $a_{j,m}$, all these individual interference components are uncorrelated. In this sense, the total ICI power can be derived through varying all the paths and all the sub-carriers.

\subsection{Total ICI Power}
To proceed, we firstly give the following lemma
\begin{lemma}
In the long term, the total ICI suffered by device $i$ is equal to the ICI incurred by device $i$ to other devices.
\end{lemma}
\begin{IEEEproof}[Proof]
From (\ref{Doppler}) and (\ref{individual-ICI-power}), the mean of $\sigma_{i,j,m}^{2}$ is given by
\begin{align}
&\bar{\sigma}_{i,j,m}^{2}=\frac{\mathbb{E}\left[\left|a_{j,m}\right|^{2}\right]P_{\textrm{T}}}{2\pi V_{\textrm{max}}}\times\\
&\int_{0}^{2\pi}\int_{0}^{V_{\textrm{max}}}\textrm{sinc}^{2}\left( \left( f_{j}-f_{i}+\frac{v}{\textrm{c}}f_{\textrm{c}}\cos( \psi ) \right)T_{\textrm{s}} \right)dv d\psi.
\label{average-individual-ICI-power}
\end{align}
Summing over all the paths, we have
\begin{equation}
\bar{\sigma}_{i,j}^{2}=\frac{P_{\textrm{T}}\int_{0}^{2\pi}\int_{0}^{V_{\textrm{max}}}\textrm{sinc}^{2}\left( \left( f_{j}-f_{i}+\frac{v}{\textrm{c}}f_{\textrm{c}}\cos( \psi ) \right)T_{\textrm{s}} \right)dv d\psi}{2\pi V_{\textrm{max}}},
\label{average-psum-ICI-power}
\end{equation}
where $\bar{\sigma}_{i,j}^{2}$ represents the average ICI power brought by device $j$ to device $i$.

Based on (\ref{average-psum-ICI-power}), it can be inferred that $\bar{\sigma}_{i,j}^{2}=\bar{\sigma}_{j,i}^{2}$, which indicates that the ICI from device $j$ to $i$ has the same average power with that from $i$ to $j$. Therefore,
\begin{equation}
P_{\textrm{ICI},i}\triangleq\bar{\sigma}_{i}^{2}= \sum_{\substack{j=-N\\j\neq i}}^{N}\bar{\sigma}_{i,j}^{2}=\sum_{\substack{j=-N\\j\neq i}}^{N}\bar{\sigma}_{j,i}^{2},
\label{average-total-ICI-power}
\end{equation}
where $P_{\textrm{ICI},i}$ is the average total ICI power suffered by device $i$. Note that $\sum_{j\neq i}\bar{\sigma}_{j,i}^{2}$ denotes the ICI power brought by device $i$ to other devices, the proof is complete.
\end{IEEEproof}

Next, to further analyze the interference power incurred by device $i$, we make the following definition.
\begin{definition}
The power leakage function $L_{i}\left( f \right)$ of device $i$ is defined as
\begin{equation}
L_{i}\left( f \right)\triangleq\frac{\int_{0}^{2\pi}\int_{0}^{V_{\emph{max}}}\emph{sinc}^{2}\left( \left( f_{i}-f+\frac{v}{\emph{c}}f_{\emph{c}}\cos( \psi ) \right)T_{\emph{s}} \right)dv d\psi}{2\pi V_{\emph{max}}}.
\label{power-leakage-function}
\end{equation}
\end{definition}
\emph{Remarks}: This function shows how much power of the signal $x_{i}$ is leaked onto the frequency band around $f$ (normalized by $P_{\textrm{T}}$). Intuitively, this function tends to be smaller when $f$ becomes larger.

Invoking (\ref{Demodulated_Signal}), (\ref{average-psum-ICI-power}) and (\ref{average-total-ICI-power}), it can be easily derived that
\begin{align}
&P_{\textrm{U},i}\triangleq \bar{\sigma}_{i,i}^{2} = L_{i}\left( f_{i} \right)P_{\textrm{T}}\label{using-power-leakage-function1},\\
&P_{\textrm{L},i}\triangleq P_{\textrm{T}}-\bar{\sigma}_{i,i}^{2} = \left(1-L_{i}\left( f_{i} \right)\right)P_{\textrm{T}}\label{using-power-leakage-function2},\\
&P_{\textrm{ICI},i}=\sum_{\substack{j=-N\\j\neq i}}^{N}L_{i}\left( f_{j} \right)P_{\textrm{T}}\label{using-power-leakage-function3}.
\end{align}
where $P_{\textrm{U},i}$ is, with an abuse of terminology, referred to as the effective useful power of signal $x_{i}$. Then it is clear that $P_{\textrm{L},i}=P_{\textrm{T}}-P_{\textrm{U},i}$ represents the leakage power of signal $x_{i}$.

From the following theorem, we know the index $i$ in $P_{\textrm{U},i}$ and $P_{\textrm{L},i}$ can be dropped.
\begin{theorem}
The effective useful power of each transmitted signal is given by
\begin{equation}
P_{\emph{U}}=\frac{2 P_{\emph{T}}}{\pi}\int_{0}^{\frac{\pi}{2}}\left[ \frac{\emph{Si}\left(2b\cos\left(\psi \right ) \right )}{b\cos\left(\psi \right )}-\frac{\sin^{2}\left(b\cos\left(\psi \right ) \right )}{\left(b\cos\left(\psi \right ) \right )^{2}} \right ]d\psi,
\label{effective-power-pro}
\end{equation}
where $b= \frac{\pi V_{\emph{max}}f_{\emph{c}}T_{\emph{s}}}{\emph{c}}$ and $\emph{Si}\left(x\right)=\int_{0}^{x}\frac{\sin\left(t\right)}{t}dt$.
\end{theorem}
\begin{IEEEproof}
Based on (\ref{power-leakage-function}), we have
\begin{align}
&L_{i}\left( f_{i} \right)\notag\\
&=\frac{1}{2\pi V_{\textrm{max}}}\int_{0}^{2\pi}\int_{0}^{V_{\textrm{max}}}\textrm{sinc}^{2}\left( \frac{f_{\textrm{c}}T_{\textrm{s}}}{\textrm{c}}v\cos( \psi ) \right)dv d\psi \notag\\
&=\frac{1}{2\pi}\int_{0}^{2\pi}\frac{\int_{0}^{\frac{\pi V_{\textrm{max}} f_{\textrm{c}}T_{\textrm{s}}}{\textrm{c}}\cos\left(\psi\right)}\textrm{sinc}^{2}\left( \frac{t}{\pi} \right)dt}{\frac{\pi V_{\textrm{max}} f_{\textrm{c}}T_{\textrm{s}}}{\textrm{c}}\cos\left(\psi\right) } d\psi \notag\\
&=\frac{2}{\pi}\int_{0}^{\frac{\pi}{2}}\frac{1}{b\cos\left(\psi\right)}\int_{0}^{b\cos\left(\psi\right)}\textrm{sinc}^{2}\left( \frac{t}{\pi} \right)dt d\psi \notag\\
&=\frac{2}{\pi}\int_{0}^{\frac{\pi}{2}}\left[ \frac{\textrm{Si}\left(2b\cos\left(\psi \right ) \right )}{b\cos\left(\psi \right )}-\frac{\sin^{2}\left(b\cos\left(\psi \right ) \right )}{\left(b\cos\left(\psi \right ) \right )^{2}} \right ]d\psi,
\label{proof-theorem-1}
\end{align}
where $b= \frac{\pi V_{\textrm{max}}f_{\textrm{c}}T_{\textrm{s}}}{\textrm{c}}$ and $\textrm{Si}\left(x\right)=\int_{0}^{x}\frac{\sin\left(t\right)}{t}dt$. The last equation is derived using the fact that $\int_{0}^{x}\textrm{sinc}^2\left( \frac{t}{\pi} \right)dt=\textrm{Si}\left(2 x\right)-\frac{\sin^{2}\left(x\right)}{x}$. Substituting (\ref{proof-theorem-1}) into (\ref{using-power-leakage-function1}) completes the proof.
\end{IEEEproof}
\emph{Remarks}: When all the devices are immobile, the effective useful power $P_{\textrm{U}}$ is equal to the transmitted power $P_{\textrm{T}}$. However, if these devices are moving, the effective useful power decreases. This theorem indicates that devices' mobility incurs a power penalty in NB-OFDMA system.

Without loss of generality, let us set the sub-carrier spacing $\Delta f$ to be $\frac{1}{T_{\textrm{s}}}$. Moreover, assume that $N$ is sufficiently large, i.e., the number of the sub-carriers $2N+1\rightarrow \infty$. Then the following lemma is derived.
\begin{lemma}
When $T_{\emph{s}}\Delta f=1$ and $N\rightarrow \infty$,
\begin{equation}
\lim_{N\rightarrow \infty}\sum_{j=-N}^{N}L_{i}\left( f_{j} \right)=1.
\label{sum-1}
\end{equation}
\end{lemma}
\begin{IEEEproof}
\begin{align}
&\lim_{N\rightarrow \infty}\sum_{j=-N}^{N}L_{i}\left( f_{j} \right)=\sum_{j=-\infty}^{\infty}L_{i}\left( f_{i}+\left(j-i\right)\Delta f \right) \notag \\
&=\sum_{j=-\infty}^{\infty}\frac{\int_{0}^{2\pi}\int_{0}^{V_{\textrm{max}}}\textrm{sinc}^{2} \left( \left( i-j \right) +\frac{ f_{\textrm{c}}T_{\textrm{s}}}{\textrm{c}}v\cos( \psi ) \right) dv d\psi}{2\pi V_{\textrm{max}}} \notag\\
&=\frac{\int_{0}^{2\pi}\int_{0}^{V_{\textrm{max}}}\sum_{k=-\infty}^{\infty}\textrm{sinc}^{2}\left( k +\frac{f_{\textrm{c}}T_{\textrm{s}}}{\textrm{c}}v \cos( \psi ) \right)dv d\psi}{2\pi V_{\textrm{max}}} \notag \\
&=\frac{\int_{0}^{2\pi}\int_{0}^{V_{\textrm{max}}}dv d\psi}{2\pi V_{\textrm{max}}}=1 .
\label{proof-lemma-2}
\end{align}
\end{IEEEproof}
\emph{Remarks}: $\Delta f=\frac{1}{T_{\textrm{s}}}$ is the minimum sub-carrier spacing to preserve the ICI-free condition in low-mobility system. Under this setting, all the leakage power become the interference to other devices. However, if $\Delta f=\frac{k}{T_{\textrm{s}}},~k\geq 2$, the interference power will be less than the leakage power. This is rather intuitive, since larger carrier spacing reduces the impact of the devices' mobility and induces smaller ICI.

Keeping (\ref{using-power-leakage-function2}) and (\ref{using-power-leakage-function3}) in mind, this lemma shows that
\begin{equation}
P_{\textrm{ICI},i}=P_{\textrm{L}}.
\label{P_ICI-P_L}
\end{equation}
Then the following theorem becomes straightforward.
\begin{theorem}
The total ICI power suffered by each device is
\begin{equation}
P_{\emph{ICI}}=P_{\emph{T}}-\frac{2 P_{\emph{T}}}{\pi}\int_{0}^{\frac{\pi}{2}}\left[ \frac{\emph{Si}\left(2b\cos\left(\psi \right ) \right )}{b\cos\left(\psi \right )}-\frac{\sin^{2}\left(b\cos\left(\psi \right) \right )}{\left(b\cos\left(\psi \right ) \right )^{2}} \right ]d\psi,
\label{average-total-ICI-power-pro}
\end{equation}
where $b=\frac{\pi V_{\emph{max}}f_{\emph{c}}}{\emph{c}\Delta f}$.
\end{theorem}
\emph{Remarks}: It can be seen that when all the devices are immobile, there is no interference between sub-carriers. And conversely, when devices move extremely fast, the useful power $P_{\textrm{U}}\rightarrow 0$ while the ICI power $P_{\textrm{ICI}}\rightarrow P_{\textrm{T}}$. Despite these observations, the relation between the ICI power and devices' mobility is still not clear.

For more insights into the impact of $V_{\textrm{max}}$ on $P_{\textrm{ICI}}$, we appeal to the following theorem.
\begin{theorem}
The total ICI power is bounded by
\begin{equation}
\left(\frac{b^{2}}{18}-\frac{b^4}{50}\right)P_{\emph{T}} \leq P_{\emph{ICI}}\leq \left(\frac{b^{2}}{18}+\frac{b^4}{60}\right)P_{\emph{T}},
\label{ICI-power-bound}
\end{equation}
where $b=\frac{\pi V_{\emph{max}}f_{\emph{c}}}{\emph{c}\Delta f}$.
\end{theorem}
\begin{IEEEproof}
It can be easily derived that for $x\in \mathbb{R}$,
\begin{equation}
x-\frac{x^3}{3!} \leq \sin(x) \leq x-\frac{x^3}{3!}+\frac{x^5}{5!},
\label{sin-bounds}
\end{equation}
\begin{equation}
\frac{x^2}{2!}-\frac{x^4}{4!} \leq 1-\cos(x) \leq \frac{x^2}{2!}-\frac{x^4}{4!}+\frac{x^6}{6!}.
\label{cos-bounds}
\end{equation}

Let us define
\begin{align}
f(\psi)&\triangleq\frac{\textrm{Si}\left( 2b\cos(\psi) \right)}{b\cos(\psi)}, \mbox{ and}\\
g(\psi)&\triangleq\frac{\sin^{2}\left(b\cos\left(\psi \right) \right )}{\left(b\cos\left(\psi \right ) \right )^{2}}=\frac{1-\cos\left( 2b\cos(\psi) \right)}{2\left(b\cos(\psi)\right)^{2}}.
\label{fg-psi}
\end{align}
Then we have
\begin{equation}
2-\frac{4}{9}\left( b\cos(\psi) \right)^{2} \leq f(\psi) \leq 2-\frac{4}{9}\left( b\cos(\psi) \right)^2+\frac{4}{75}\left( b\cos(\psi) \right)^{4},
\label{sin-bounds-pro}
\end{equation}
\begin{equation}
1-\frac{1}{3}\left( b\cos(\psi) \right)^{2} \leq g(\psi) \leq 1-\frac{1}{3}\left( b\cos(\psi) \right)^2+\frac{2}{45}\left( b\cos(\psi) \right)^{4}.
\label{cos-bounds-pro}
\end{equation}
Substituting (\ref{sin-bounds-pro}) and (\ref{cos-bounds-pro}) into (\ref{average-total-ICI-power-pro}), the proof is complete.
\end{IEEEproof}
\emph{Remarks}: The total ICI power is proportional to the effective transmission power $P_{\textrm{T}}$, which is quite intuitive. Given that $P_{\textrm{T}}$, $f_{\textrm{c}}$ and $\Delta f$ are generally fixed in a real system, $P_{\textrm{ICI}}$ can be expressed as a polynomial in $V_{\textrm{max}}^{2}$.

As shown in \cite{3gpp2015mobile}, the fundamental parameters of a conventional NB-OFDMA are: carrier center frequency $f_{\textrm{c}}=900$MHz, sub-carrier spacing $\Delta f=2.5$kHz. Without loss of generality, let us set $V_{\textrm{max}}$ as $100$m/s. Then it can be easily derived that $b=\frac{\pi V_{\textrm{max}}f_{\textrm{c}}}{\textrm{c}\Delta f}\approx 0.377$. Under this setting, we have $\frac{b^{4}}{60}<\frac{b^{4}}{50}\ll \frac{b^{2}}{18}$,\footnote{We use $\left| x \right|\ll \left| y \right|$ to represent that $\left| x \right|< \left| y \right|/10$.} thereby justifying the following corollary.
\begin{corollary}
When $b$ is small enough, the total ICI power can be approximated by
\begin{equation}
P_{\emph{ICI}}\approx \frac{1}{18}\left(\frac{\pi V_{\emph{max}}f_{\emph{c}}}{\emph{c}\Delta f} \right)^{2} P_{\emph{T}}.
\label{average-total-ICI-power-app}
\end{equation}
\end{corollary}
\emph{Remarks}: This corollary shows that the total ICI power is approximately proportional to $V_{\textrm{max}}^{2}$ in low $b$ regime. Moreover, we see higher center frequency and smaller sub-carrier spacing incur higher interference power.

A sufficient condition for $\frac{b^{4}}{50}\ll \frac{b^{2}}{18}$ is $b=\frac{\pi V_{\textrm{max}}f_{\textrm{c}}}{\textrm{c}\Delta f}<\frac{1}{2}$. Hence, it can be inferred that the approximated $P_{\textrm{ICI}}$ is rather tight as long as $V_{\textrm{max}}<\frac{\textrm{c}\Delta f}{2\pi f_{\textrm{c}}}$.

\subsection{Uplink Sum-rate}

Based on the above analysis, now we are ready to formulate the relationship between the system sum-rate and the devices' mobility.

Let us first look at the ergodic capacity of each device, which is given by
\begin{equation}
\textrm{C}_{i}=\mathbb{E}\left[\log\left(1+\frac{\sigma_{i,i}^{2}}{\sum_{j\neq i}\sigma_{i,j}^{2}+\sigma_{\textrm{n}}^{2}} \right)\right].
\label{single-rate}
\end{equation}
Invoking Jensen inequality, an upper bound of $\textrm{C}_{i}$ can be easily derived as
\begin{equation}
\textrm{C}_{\textrm{upper}}=\log\left(1+\frac{P_{\textrm{U}}}{P_{\textrm{ICI}}+\sigma_{\textrm{n}}^{2}} \right),
\label{single-rate-upper}
\end{equation}
where $P_{\textrm{U}}$ and $P_{\textrm{ICI}}$ are determined invoking Theorem 1 and Theorem 2, respectively.

The following theorem provides a direct relationship between the ergodic capacity and devices' mobility.
\begin{theorem}
The upper bound of ergodic capacity can be approximated by
\begin{equation}
\emph{C}_{\emph{upper}}\approx -\log\left( \frac{1}{18}\left(\frac{\pi V_{\emph{max}}f_{\emph{c}}}{\emph{c}\Delta f}\right)^{2}+\frac{\sigma_{\emph{n}}^{2}}{P_{\emph{T}}}\right) +\frac{\sigma_{\emph{n}}^{2}}{P_{\emph{T}}}.
\label{single-rate-pro}
\end{equation}
\end{theorem}
\begin{IEEEproof}
We know that $b$ is small enough in a real system, so it is reasonable to obtain $P_{\textrm{ICI}}$ using Corollary 1. Therefore, (\ref{single-rate}) can be rewritten as
\begin{align}
\textrm{C}_{\textrm{upper}}&\approx \log\left(1+\frac{1-\frac{1}{18}b^{2} }{\frac{1}{18}b^{2}+\frac{\sigma_{\textrm{n}}^{2}}{P_{\textrm{T}}}} \right) \notag\\
          &\approx \log\left(\frac{1-\frac{1}{18}b^{2} }{\frac{1}{18}b^{2}+\frac{\sigma_{\textrm{n}}^{2}}{P_{\textrm{T}}}}\right)+\frac{\frac{1}{18}b^{2}+\frac{\sigma_{\textrm{n}}^{2}}{P_{\textrm{T}}}}{1-\frac{1}{18}b^{2}} \notag\\
          &\approx -\log\left( \frac{1}{18}b^{2}+\frac{\sigma_{\textrm{n}}^{2}}{P_{\textrm{T}}}\right) +\frac{\sigma_{\textrm{n}}^{2}}{P_{\textrm{T}}},
\label{single-rate-proof}
\end{align}
which completes the proof.
\end{IEEEproof}
\emph{Remarks}: It is rather intuitive that the ergodic capacity tends to be smaller when $V_{\textrm{max}}$ increases. Moreover, as SNR grows high, an insightful observation is that $\textrm{C}_\textrm{upper}$ decreases linearly with $V_{\textrm{max}}$ on a logarithmic scale. In the meanwhile, we see higher center frequency and smaller sub-carrier spacing also incur lower capacity.

Now the upper bound of the uplink sum-rate becomes quite straightforward, which is
\begin{equation}
\textrm{C}_{\textrm{sum,upper}}=\textrm{B}\textrm{C}_{\textrm{upper}},
\label{sum-rate}
\end{equation}
where $\textrm{B}$ is uplink system bandwidth, $\textrm{C}_{\textrm{upper}}$ is derived through Theorem $4$. Combined with Theorem $4$, it is clear that the $\Delta f$ and other involved parameters must be carefully designed to suit the differentiated service provisioning and dynamic number of users in a mobile IoT networks.

\section{Numerical Results}

In this section, we study the performance of a mobile NB-OFDMA system through a serious of Monte Carlo simulations. We first verify the relationship between the total ICI power and devices' maximum velocity. Subsequently, the ergodic capacity and system sum-rate induced by distinct carrier spacings and different SNR conditions are investigated. For all simulation examples, we set $P_{\textrm{T}}=1$ for clarity purpose. The wave propagation speed is selected as $\textrm{c}=3\times 10^{8}$m/s.

\begin{figure}[!ht]\centering
\includegraphics[angle=0,scale=0.42]{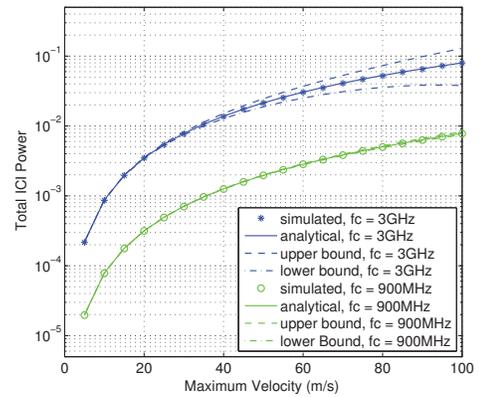}
\caption{Total ICI power for $f_{\textrm{c}}=3$GHz, $900$MHz.}
\label{Fig_ICI}
\end{figure}

In the first example, we compare the simulated and analytical ICI as devices' maximum velocity increases. We set the sub-carrier spacing as $\Delta f=2.5$KHz and choose the carrier center frequency $f_{\textrm{c}}$ as $3$GHz and $900$MHz. Fig. \ref{Fig_ICI} shows that the analytical result coincides with the simulated values, thereby verifying Theorem $2$. Moreover, it can be seen that higher velocity and higher center frequency lead to more serious ICI. As for the upper and lower bounds, they are tight for $f_{\textrm{c}}=900$MHz. Yet when $f_{\textrm{c}}=3$GHz, these bounds are tight only for $V_\textrm{max}<40$m/s. The reason is that $V_{\textrm{max}}<\frac{\textrm{c}\Delta f}{2\pi f_{\textrm{c}}}\approx 40$m/s is required when invoking Theorem $3$. Bases on the above observations, Theorem $3$ is also validated.
\begin{figure}[!ht]\centering
\includegraphics[angle=0,scale=0.42]{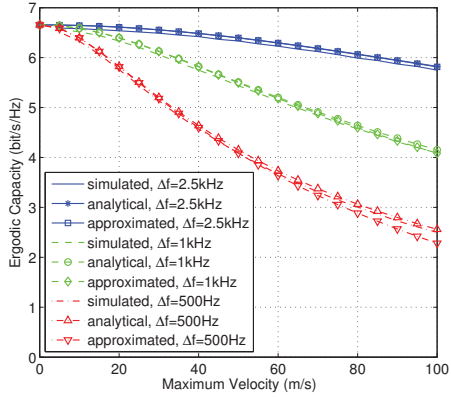}
\caption{Ergodic capacity for SNR$=20$dB, $\Delta f=2.5$KHz, $1$KHz and $500$Hz.}
\label{Fig_ergodicC}
\end{figure}

\begin{figure}[!ht]\centering
\includegraphics[angle=0,scale=0.42]{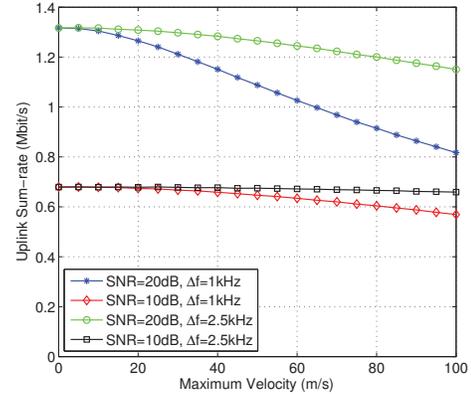}
\caption{Uplink sum-rate for $\Delta f=2.5$KHz, $1$KHz under different SNR conditions.}
\label{Fig_sumC}
\end{figure}

In the next example, we investigate the impacts of sub-carrier spacing and SNR on the ergodic capacity of each device and uplink sum-rate with $f_{\textrm{c}}=900$MHz. Fig. \ref{Fig_ergodicC} shows the analytical and approximated ergodic capacity upper bound using (\ref{single-rate}) and (\ref{single-rate-pro}), respectively, for $\Delta f=2.5$KHz, $1$KHz and $500$Hz. It is intuitive that higher velocity and smaller sub-carrier spacing result in smaller ergodic capacity. We see the approximated result is quite close to the exact upper bound. In the meantime, the upper bound of the uplink sum-rate is depicted in Fig. \ref{Fig_sumC}. Since the NB-OFDMA system is usually deployed in a system bandwidth that is multiple of $200$KHz, we choose $B=200$KHz for simplicity. It can be observed from Fig. \ref{Fig_sumC} that lower velocity and larger sub-carrier spacing induce higher uplink sum-rate, at the price of decreasing the number of devices, though. Moreover, it is rather intuitive that $V_{\textrm{max}}$ has more significant effect on the system sum-rate when SNR is high, since it can be inferred from (\ref{single-rate}) and (\ref{single-rate-pro}) that bigger $\sigma_{\textrm{n}}^{2}$ downplays the impact of interference counterpart $P_{\textrm{ICI}}$.

\section{Conclusion}
In this paper, we first model the devices' mobility and multi-path scattering environments using a series of Possion point processes. Then, through summarizing the individual ICI arising from each path of all the other devices, we obtain the average total ICI power as a function of devices' maximum velocity, carrier frequency and sub-carrier spacing. On these basis, we heuristically formulate the relationship between the system sum-rate and devices' mobility. These results provide reference for the system design in future mobile IoT networks.

\section{Acknowledgement}
This work was supported in part by National Key Basic Research Program of China (No. 2012CB316104), National Hi-Tech R\&D Program (No.2014AA01A702), National Natural Science Foundation of China (No. 61371094), Zhejiang Provincial Natural Science Foundation of China (No.LR12F01002).

\bibliographystyle{IEEEtran}

\end{document}